\documentclass[twocolumn,showpacs,preprintnumbers]{emulateapj}%
\usepackage{amsfonts}
\usepackage{amsmath,epsfig}
\usepackage{graphicx}
\usepackage{dcolumn}
\usepackage{bm}
\usepackage{amssymb}
\usepackage{amsmath}%
\begin{document}

\title{Merger of two neutron stars: predictions from the two-families scenario}
\author{Alessandro Drago$^{\text{(a)}}$ and
Giuseppe Pagliara$^{\text{(a)}}$ }
\affiliation{$^{\text{(a)}}$Dip.~di Fisica e Scienze della Terra dell'Universit\`a di Ferrara and INFN
Sez.~di
Ferrara, Via Saragat 1, I-44100 Ferrara, Italy}

\begin{abstract}
If only one family of "neutron stars" exists, their maximum mass must
be equal or larger than $2 M_\odot$ and then only in less than about $18\%$ of cases the outcome of
the merger of two neutron stars is a prompt collapse to a black
hole, since the newly formed system can avoid the collapse at
least until differential rotation is present. In the so-called
two-families scenario, stars made of hadrons are stable only up to
about $(1.5-1.6) M_\odot$, while the most massive compact stars are entirely made
of strange quark matter. We show that in this scenario the outcome of
the merger of two compact stars, entirely composed by hadrons, is a prompt collapse in at least $34\%$ of the
cases. It will therefore be easy to discriminate between the two
scenarios once the gravitational waves emitted at the moment of the
merger are detected. Finally, we shortly discuss the implications of GW170817-GRB170817A. 
\end{abstract}   
\maketitle

The detection of Gravitational Waves (GWs) has made available a new tool
to investigate the properties of matter at extreme density.
In particular, the future detection of GWs from neutron star - neutron star
mergers will provide information about the Equation of State (EoS) of matter
from the analysis of both the inspiral and of the postmerger
phase \citep{Baiotti:2016qnr}. One of the main open questions concerns
the composition of matter at the center of a compact star:
hyperons, delta resonances or even deconfined quark matter
could appear. Quark matter could be present only at the center
of the star (hybrid stars) or occupy the whole star (Strange Quark Stars, SQSs)
\citep{Bodmer:1971we,Witten:1984rs,Alcock:1986hz,Haensel:1986qb}.

What characterizes the two-families scenario is the idea
that by increasing the central density of a compact star,
more and more resonances are produced (deltas, hyperons etc.)
and this (in the absence of quark deconfinement) leads to a dramatic softening of the EoS
entailing a small value of the maximum mass of Hadronic Stars, HSs, (smaller than about $2M_{\odot}$)
and the possibility of very small radii. If quark deconfinement 
can take place, the EoS becomes much stiffer and 
stable configurations with masses up to $2M_{\odot}$ (or more) can be 
obtained as SQSs. 
Thus in the two-families scenario, HSs and SQSs coexist
\citep{Drago:2015cea,Drago:2015dea}. HSs can be very compact (with
radii smaller than about $11$ km) and have a maximum mass
$M_{\mathrm{max}}^H \sim (1.5 - 1.6) M_{\odot}$. The small radius of
these stellar objects is mainly due to the appearance of delta
resonances. On the other hand, SQSs are larger and can reach a maximum
mass $M_{\mathrm{max}}^Q$ which in principle can be even significantly larger than
$2M_{\odot}$ \citep{Kurkela:2009gj,Fraga:2013qra}.

The co-existence of these two families
implies three possible types of mergers:
HS-HS, HS-SQS and SQS-SQS. In the present letter we concentrate on
the first possibility.

The two families scenario is based on the so called Bodmer-Witten hypothesis
\citep{Bodmer:1971we,Witten:1984rs}
for which the true ground state of strongly interacting matter
is not nuclear matter but strange quark matter.
In this scenario, strange quark matter would appear not only in
stellar size objects but also in ``small'' nuggets, named strangelets,
which could be formed for instance during the merger of two SQSs \citep{Bauswein:2008gx}.
Strangelets propagating within the galaxy could in principle trigger the conversion of all neutron stars
into SQSs \citep{Madsen:1989pg} but it has been shown in 
\citet{Wiktorowicz:2017swq} that the galactic density of strangelets due to SQS mergers can be as small as
$10^{-35}$ gr/cm$^3$.
For the present discussion, we assume that the process of conversion of a HS into a SQS
is never triggered by external seeding from a strangelet.
Instead an HS can convert ``spontaneously'' into SQS once a sizable
  fraction of strangeness appears in its core via hyperons' formation.
 The formation of hyperons can be due to the increase of the central density of the
  star, originated e.g. by the magnetic driven spin down in the case of an
  isolated neutron star \citep{Pili:2016hqo}, or by mass accretion in binary systems 
  \citep{Wiktorowicz:2017swq}. 
The process of deconfinement starts when a critical density is reached,
  which corresponds (for cold and non rotating stars) to a critical mass slightly smaller than $M_{\mathrm{max}}^H$. Rotation and temperature
  can modify the value of the critical mass.

  The process of conversion can be divided into two different stages \citep{Drago:2015fpa}:\\
  a) a turbulent combustion which, in a time scale $t_{\mathrm{turb}}$ of the order of a few ms, converts most of the 
  star; \\
  b) a diffusive combustion which converts the unburnt hadronic layer in a time scale $t_{\mathrm{diff}}$ of the order of ten seconds.
  It has been shown that the hybrid star configuration, HybS, obtained after phase a), is roughly as stiff as the final SQS configuration and therefore
  has a maximum mass $M_{\mathrm{max}}^{Hyb} \sim M_{\mathrm{max}}^Q$.
  
  The merger of two neutron stars could possibly lead to the formation of a SQS \citep{Cheng:1995am,Drago:2015qwa}. 
  A necessary
  condition is that the newly formed system lives at least for a time scale of the order of $\sim t_{\mathrm{turb}}$.
  A prompt collapse occurs if even a strong differential rotation
  is not able to delay the collapse to a Black Hole (BH). When this happens,
  the collapse takes place within $t_{\mathrm{coll}} \sim 1$ms. Therefore in the case of a prompt collapse,
  $t_{\mathrm{coll}} \sim 1 \mathrm{ms} <  t_{\mathrm{turb}}$ and quark deconfinement does not even start. The only relevant
  EoS in this case is the hadronic one.

The condition for obtaining a prompt collapse is that the mass of the
newly formed compact object exceeds the maximum mass of hypermassive
stars $M_{\mathrm{max,dr}}$. A first simple estimate of $M_{\mathrm{max,dr}}$ can be done 
by using the very recent analysis of \citet{Weih:2017mcw}:
by using several zero temperature EoSs and by adopting the commonly
used constant angular momentum law, it has been found, within a full
GR code, that $M_{\mathrm{max,dr}} = (1.54 \pm 0.05) M_{\mathrm{TOV}}$
with $M_{\mathrm{TOV}}$ being the maximum mass of cold and
non-rotating stellar configurations. By setting $M_{\mathrm{max}}^H=1.6 M_{\odot}$,
one obtains $M_{\mathrm{max,dr}} \sim 2.5 M_{\odot}$ within the two families
scenario and $M_{\mathrm{max,dr}} \sim 3 M_{\odot}$ within the one family scenario (assuming a maximum mass of $2M_{\odot}$).
Notice however that the real angular velocity profile after the merger 
can only be obtained through explicit simulations of the merger as done
in \citet{Bauswein:2013jpa,Bauswein:2015vxa,Bauswein:2017aur} within the conformal flatness approximation of Einstein’s field equations.
In these studies also thermal effects have been included
by using tabulated finite temperature EoSs. Notice that the thermal pressure
helps in stabilizing the remnant.
It turns out that $M_{\mathrm{threshold}}$ (the maximum mass not leading to a prompt collapse)
depends on the compactness
of the merging stars and it can be as high as $1.7M_{\mathrm{TOV}}$.
Finally, general relativistic hydrodynamics simulations of the merger in a full GR framework have been performed in 
\citet{Hotokezaka:2013iia,Feo:2016cbs,Maione:2017aux}
for a few representative EoSs and by parametrizing the thermal effects with an effective adiabatic index.

Independently from the actual value of $M_{\mathrm{max,dr}}$,
the key point is that in the case of a prompt collapse, within the two-families scenario, $M_{\mathrm{TOV}}$ corresponds to $M_{\mathrm{max}}^H$
  and not to $M_{\mathrm{max}}^Q$ because there is not enough time for the SQS to 
  start forming at the center of the newly born stellar object.
As we will explain in the following, this difference
  between the one-family and the two-families scenario will allow to unambiguously rule out one of them already after a few
detections of Gravitational Waves (GWs) by the LIGO and VIRGO experiments.

We can make predictions on the fate of a merger by using the present knowledge
on the mass distribution of compact stars in binary systems. In \citet{Kiziltan:2013oja}, it is shown that the mass
distribution of pulsars in double neutron star systems peaks at
$1.33 M_{\odot}$ with a $\sigma \sim 0.11M_{\odot}$. Thus, by assuming that this distribution
coincides with the mass distribution of all neutron stars in binary systems, 
one can
estimate the distribution of the total mass of merging binaries $M_{\mathrm{tot}}$ as
peaked at $2.66 M_{\odot}$ with $\sigma \sim \sqrt{2}\times0.11M_{\odot}$.
Notice however that the mass distribution of systems merging within a Hubble time
could be shifted to larger values with respect to the distribution of \citet{Kiziltan:2013oja}.

We can now estimate the fraction of events which lead to a prompt collapse as follows.
We adopt the empirical relations for $M_{\mathrm{threshold}}$ which have 
been obtained by fitting the results of explicit numerical simulations of mergers \citep{Bauswein:2013jpa,Bauswein:2015vxa,Bauswein:2017aur}.
In particular, 
we use the relation between $M_{\mathrm{threshold}}$ and the compactness of the maximum mass configuration, $C_{\mathrm{max}}=M_{\mathrm{TOV}}/R_{\mathrm{TOV}}$
which reads: $M_{\mathrm{threshold}}=(2.43-3.38\times C_{\mathrm{max}})\times M_{\mathrm{TOV}}$.
By using this parametrization, within the one-family scenario, one can notice from Table 1 of \citet{Bauswein:2017aur}
that the minimum value of $M_{\mathrm{threshold}}$ is of the order of $2.8M_{\odot}$.
This result has been obtained by many independent simulations.
In \citet{Hotokezaka:2013iia}, among the six different EoSs used for the numerical simulations, the SLy EoS \citep{Douchin:2001sv}
provides the smallest value of $M_{\mathrm{threshold}}$ which turns out to be of the order of $2.8 M_{\odot}$.
Similar results for the Sly EoS have been obtained within the numerical simulations of \citet{Feo:2016cbs,Maione:2017aux}.
By using the mass distribution of \citet{Kiziltan:2013oja}, the probability of a prompt collapse $\mathrm{P_{prompt}}$ 
turns out to be $\mathrm{P_{prompt}}<18\%$ (see lower panel of Fig.1). We regards this number as an upper limit for the rate of prompt collapses
within the one family scenario.

For the two-families scenario, by varying $M_{\mathrm{max}}^H$ in the range $(1.5-1.6) M_{\odot}$ and the corresponding
radii within the range $(10-11)$km we can compute $C_{\mathrm{max}}$ and thus $M_{\mathrm{threshold}}$ which turns
out to vary in the range $(2.52-2.72) M_{\odot}$. Correspondingly,  $34\%<\mathrm{P_{prompt}}<82\%$ (see upper panel of Fig.1).
It is clear therefore that within
the two-families scenario one expects a significant number of prompt collapses whereas within the one-family scenario
this possibility is suppressed.

From all these analyses one can conclude that the two-families scenario predicts a number of prompt collapses 
significantly larger than in the case of the one-family scenario.
Therefore in the near future it will be possible to rule out one of the two scenarios.
Indeed, the signal emitted in the case of a prompt
collapse is clearly distinguishable from the signal of a
differentially rotating remnant, see \citet{Baiotti:2016qnr} for a recent review.

The cases (in both scenarios) in which the post-merger remnant is stable, for at
least a few ms, deserve a separate
discussion. Remarkably, the GW signal emitted from the remnant can also
bring important information on the EoS. 
There are several studies indicating that
a Fourier analysis of the post-merger GW signal allows to identify the predominant
oscillation mode, whose frequency (indicated with $f_{peak}$ in
\citet{Bauswein:2011tp} and $f_2$ in \citet{Stergioulas:2011gd,Takami:2014zpa,Maione:2017aux}) 
depends strongly on the stiffness of the EoS: stiffer EoSs predict smaller values of $f_2$.
Moreover, sub-dominant modes at frequencies lower than the one of $f_2$ have been identified 
in \citet{Stergioulas:2011gd,Takami:2014zpa,Bauswein:2015yca,Maione:2017aux} which, if detected together with the $f_2$ mode,
could strongly constrain the EoS.

Let us now discuss which are the expected signatures in the GWs signal
of the two-families scenario during the postmerger phase. In this
scenario, the postmerger remnant is at the beginning very compact
(because the star is made of hadronic matter). We can estimate the
initial value of $f_2$ by using the empirical relation found in
\citet{Bauswein:2015vxa}: $f_2$ ranges from $3.3$ to $3.7$ kHz for
$M_{\mathrm{tot}}$ ranging from $2.4$ to $2.7 M_{\odot}$. Here we have
assumed the radius of the $1.6M_{\odot}$ star to be
$R_{1.6} \sim 11$km.  Notice that also some purely nucleonic EoSs, such as APR4 and Sly,
predict very large values of $f_{2}$
\citep{Bauswein:2011tp,Takami:2014zpa,Maione:2017aux}. However the expected number
of events of prompt collapse in those cases would be
significantly smaller respect to the one predicted in the two-families
scenario. Moreover, two major drawbacks of these type of EoSs must be remarked: 
first, the center of the compact star
reaches densities so high (see \citet{Hanauske:2016gia}) that it seems
unrealistic to neglect non-nucleonic degrees of freedom. Secondly, they
predict a radius for the  $1.4M_{\odot}$ configuration smaller than $12$km.
On the other hand, the recent meta-modeling analysis of \citet{Margueron:2017lup}, which is
based only on nuclear physics constraints, has suggested that compact stars composed exclusively of nucleons and leptons
have a radius of $12.5\pm 0.4$ km for masses ranging from $1$ up to $1.6 M_{\odot}$.

A second feature of the two-families scenario is linked to the moment
in which quarks start being formed in the center of the compact star.
Once the burning process is triggered, the stiffening of the EoS,
resulting from the formation of quark matter, leads to a significant
structural change of the central part of the star within a time scale
of the order of a few ms \citep{Pagliara:2013tza,Drago:2015fpa}.
Consequently, also the spectrum of the emitted GWs should be
significantly different with respect to the one displayed during the
first milliseconds.  It is very difficult though to make even
qualitative predictions on such a modification of the spectrum because
there are at least two different mechanisms potentially shifting $f_2$
to opposite directions. The stiffening of the EoS entails a larger
radius (at least for a non-rotating star) but at the same time it
increases the moment of inertia thus reducing the rotational frequency
of the star.  Since the equatorial radius increases with the
rotational frequency, those two effects could potentially compensate
when studying deconfinement in a rapidly rotating star.  In
conclusion, while the process of deconfinement will surely leave an
imprint on the spectrum, it is not clear if $f_2$ will significantly
change and in which direction.

The fate of compact star mergers is related also to the
phenomenology of Short-Gamma-Ray-Bursts SGRBs \citep{Gao:2015xle,Piro:2017zec}.
In particular, the two mostly discussed models for their inner
engine are based either on the formation of a
rapidly spinning BH surrounded by a hot and highly magnetized torus
\citep{Rezzolla:2011da}
or on the formation of a protomagnetar \citep{Duncan:1992hi,Rowlinson:2013ue}.
One can roughly divide SGRBs in two sub-classes: those displaying only a prompt emission
and typically lasting a fraction of a second and those in which some form of prolonged
emission is present. The existence of an extended X-ray emission
can be modelled by assuming the formation of a
supramassive and highly magnetized star \citep{Rezzolla:2014nva,Ciolfi:2014yla,Lu:2015rta,Drago:2015qwa}
while hypermassive stars are associated with SGRBs displaying only a prompt emission.
Within the one-family scenario, in \citet{Piro:2017zec} it has been shown
that to populate both sub-classes $M_{\mathrm{TOV}}$
should be close to $2M_{\odot}$ (for larger values of $M_{\mathrm{TOV}}$ 
one needs to assume that a significant fraction of SGRBs are due to BH-NS mergers).
Similar considerations apply also to the two-families scenario.
For instance, let us set $M_{\mathrm{max}}^H=1.6 M_{\odot}$, $M_{\mathrm{max}}^Q\sim M_{\mathrm{max}}^{Hyb}=2M_{\odot}$ and 
let us assume  $M_{\mathrm{threshold}}=1.6\times M_{\mathrm{max}}^H=2.56 M_{\odot}$

By adopting for the maximum mass of supramassive HybS $M_{\mathrm{supra}}^{Hyb}=1.2\times M_{\mathrm{max}}^{Hyb}=2.4M_{\odot}$
\citep{Breu:2016ufb}, we obtain
the following possible outcomes for the merger depending on its gravitational mass $M_g$:\\
a) if $M_g>M_{\mathrm{threshold}}=2.56 M_{\odot}$ we have a direct collapse to a BH without any significant prompt emission;\\
b) if $M_{\mathrm{supra}}^{Hyb}=2.4M_{\odot}<M_g<M_{\mathrm{threshold}}=2.56 M_{\odot}$ we have the formation of a hypermassive HyBS (SGRBs without extended
emission);\\
c)if $M_g<M_{\mathrm{supra}}^{Hyb}=2.4M_{\odot}$ we have the formation of supramassive HyBS which can be associated with
SGRBs with an extended emission.

To estimate the fractions of mergers populating these three cases one needs to
compute the relation between gravitational mass and baryonic mass for rapidly rotating
stars, as done in \citet{Piro:2017zec}. From a qualitative viewpoint,
we can conclude that also in the two-families scenario if $M_{\mathrm{max}}^Q \sim 2M_{\odot}$ 
it is possible at least in principle to assume that all SGRB are due to NS-NS merger.\footnote{It is interesting to notice that, while values of $M_{\mathrm{max}}^Q$ significantly larger than $\sim 2M_{\odot}$ 
have been discussed in the literature \citep{Kurkela:2009gj,Fraga:2013qra}, 
when chiral models are used to describe the quark dynamics
the value of $M_{\mathrm{max}}^Q$ is $\sim 2M_{\odot}$ \citep{Chen:2016ran,Dondi:2016yjl}.}

Two models have been proposed to explain the extended emission, both
based on the formation of a supramassive star: in
\citet{Rezzolla:2014nva,Ciolfi:2014yla} the prompt emission is
produced by the collapse of the supramassive star to a BH, due to the
magnetic spin-down having a time scale which can easily exceed $10^3$
s.  In this model one needs to assume a ``time-reversal'' scenario in
which the extended emission is generated before the prompt emission
but it appears later. In \citet{Drago:2015qwa}, the prompt emission
is due to the formation of a SQS and the extended emission is powered
by the supramassive SQS: no time reversal is needed in this case. The
delay between the merger and the prompt is due to the time needed to
convert completely the HyBS into a SQS and it is of the order of $10$ s.  The
detection of both the GW signal at merger and the electromagnetic
emission of the prompt will allow to easily distinguish between the
two models.

\begin{figure}[ptb]
\vskip 1cm
\begin{centering}
\epsfig{file=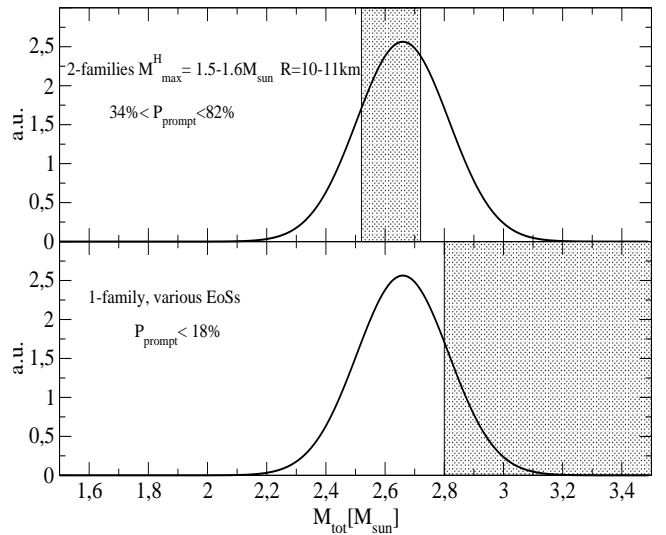,height=8.5cm,width=7cm,angle=-90}
\caption{Distribution of $M_{\mathrm{tot}}=m_1+m_2$ (solid line) estimated by using the
analysis of \citet{Kiziltan:2013oja}. In the upper panel the range of values of $M_{\mathrm{threshold}}$ for $M_{\mathrm{max}}^H=(1.5-1.6) M_{\odot}$ are indicated by the grey area (two-families scenario). In the lower panel the range of values
of $M_{\mathrm{threshold}}$ in the one-family case is indicated. In this figure the results of the analyses of
\citet{Bauswein:2013jpa,Bauswein:2015vxa,Bauswein:2017aur,Hotokezaka:2013iia,Feo:2016cbs,Maione:2017aux} have been used.}
\end{centering}
\end{figure}

The analyses presented in this paper are based on simulations which do not take into account the effect of the viscous dissipation.
Very recently the effect of shear and bulk viscosity has been investigated
in \citet{Alford:2017rxf} and \citet{Shibata:2017xht}. As shown in \citet{Shibata:2017xht},
the lifetime of the hypermassive configuration can be significantly reduced
if a rather large value of the shear viscosity is assumed. On the other hand,
since we have discussed prompt collapses occurring on a time scale of $\lesssim 1$ ms,
one can notice from \citet{Shibata:2017xht} that viscous dissipation plays a marginal role in this case.

Finally, let us summarize what one can learn from GWs detections
concerning the two-families scenario assuming that both stars are HSs.
Observations can lead to falsification and confirmation tests.\\

{\it Tests falsifying the model:}

-- No evidence of rapid collapse to a BH (within a few ms from the merger)
          for a system having total
          mass larger than $M_{\mathrm{threshold}}$, whose maximum value is of about
          $2.7M_{\odot}$.
          E.g., the merger of two $1.4 M_{\odot}$ HSs 
          would rule out the two-families scenario if it does not collapse immediately into a BH.

-- Indications, during the inspiral and/or during the first milliseconds of
the postmerger phase, of a very stiff EoS (low values of $f_{2}$, smaller than about $3$ kHz, \citep{Maione:2017aux}).\\

{\it Tests against the model although not conclusive:}

No significant change of the spectrum during the first few tens
milliseconds (the conversion to quark matter could occur at later times when the GWs signal is too weak to be detectable). \\

{\it Validating (but not conclusive) tests:}

Very low stiffness of the EoS during the inspiral or immediately after
          the merger ($f_{2}$ larger than about $3.3$ kHz).\\

{\it Strong confirmation tests:}

Rapid collapse to a BH of a merger having a total gravitational mass smaller than about $2.7 M_{\odot}$.

\vskip 0.5cm
During the process of review of this paper, the first
detection of gravitational waves from a neutron star merger has been announced: GW170817 \citep{TheLIGOScientific:2017qsa}.
Together with the GW signal, also a SGRB, GRB170817A, and a delayed kilonova, AT2017gfo, have been
detected \citep{GBM:2017lvd,Monitor:2017mdv}. From the GW signal, it has been possible to measure
the total mass of the system which turns out to be $M_{\mathrm{tot}}=2.74^{+0.04}_{-0.01} M_{\odot}$.
Although not completely excluded, the possibility that the merger led to a prompt collapse is very
unlikely since in that case it would be very difficult to explain the SGRB and the kilonova
\citep{Margalit:2017dij,Bauswein:2017vtn}. Therefore
one can conclude that  $M_{\mathrm{threshold}} > 2.74 M_{\odot}$.
As we have shown in this work, within the two families scenario $M_{\mathrm{threshold}}$ can barely reach $2.72 M_{\odot}$
in the case of a HS-HS merger. On the other hand, in a HS-SQS merger, a HybS configuration forms
without any delay and a prompt collapse is avoided. This hybrid configuration lasts $\sim 10$ s \citep{Drago:2015fpa}, long enough
to allow the emission of the nuclear material which then produces the kilonova.
Further analyses and discussions will be presented in a forthcoming paper.

We thus conclude that in our scenario GW170817 is likely not a HS-HS merger
but could be a HS-SQS merger.

\vskip 0.5cm
We would like to thank Luciano Rezzolla: a discussion with him was the origin of the present work. We would also like to thank Alessandra Feo, Roberto De Pietri and Andreas Bauswein for many useful suggestions especially in connection with
the determination of $M_{\mathrm{threshold}}$.

\vskip 0.1cm


\end{document}